\pdfoutput=1
\documentclass[12pt,preprint,epsf]{aastex} 

\usepackage{epstopdf}

\shorttitle{Orbit of Mars}
\shortauthors{Krisciunas} 

\begin{document}
\received{3 January 2019}

\title{Demonstrating the Elliptical Orbit of Mars using Naked Eye Data}   
\author{
Kevin Krisciunas\altaffilmark{1}
}
\altaffiltext{1}{George P. and Cynthia Woods Mitchell Institute for Fundamental 
Physics \& Astronomy, Texas A. \& M. University, Department of Physics \& Astronomy,
  4242 TAMU, College Station, TX 77843; {krisciunas@physics.tamu.edu} }

\begin{abstract} 

Over the course of 3.1 years we determined the 
position of Mars on 75 dates using a handheld cross staff and 
two to five bright reference stars of known right ascension and 
declination on each occasion.  On average the observed positions 
of Mars are within $\pm$11 arc minutes of the true ecliptic latitudes and 
ecliptic longitudes. After deriving the two dates of opposition to the Sun,
we were able to carry out a two stage analytical experiment on Mars' orbit. 
From the 2015/2016 data we obtain a value of the eccentricity of 
0.093 $\pm$ 0.012.  The 2015/2016 data can be fit reasonably well 
by adopting a circular orbit for the Earth, but the 2017/2018 data
must be fit with an ellipse for Mars and an ellipse for the Earth.
Applying the two ellipse model to the 2015/2016 data, we obtain an RMS error
of the ecliptic longitudes of only $\pm$7.5 arc minutes. While
Kepler was able to derive the {\em shape} of Mars' orbit while relying on
data of Tycho Brahe accurate to $\pm$2 arc minutes, today we may assume
an ellptical orbit, and we can show that much less accurate data 
are consistent with an ellipse having an eccentricity equal to the modern 
accepted value of 0.0934, within the errors.

\end{abstract}


\section{Introduction}

The fundamental paradigm of solar system astronomy prior to the time 
of Copernicus was that the Earth was at the center of the solar 
system.  Also, celestial bodies were assumed to move along perfect 
circles.  This led to the system of deferents and epicycles.  One 
prime motivation for the use of epicycles was to account for 
retrograde motion. Copernicus' great book {\em On the Revolutions of 
the Heavenly Spheres} (1543) asserted that the Sun is physically and 
truly at the center of the solar system, and that the motions of the 
planets, including the Earth, in a heliocentric system provide a 
much simpler explanation for the retrograde motion of the planets. 
However, Copernicus retained circular motion.  Also, he retained the 
notion of epicycles to account for variations of distance of the 
planets from the Sun \citep{Ging93}.

In the {\em Almagest} Ptolemy gives values for the minimum and maximum angular 
sizes of the Moon of 31\arcmin ~20\arcsec and 35\arcmin ~20\arcsec, respectively 
\citep{almagest}. Naked eye observations by this author have demonstrated that 
one can show, without using a telescope, that the angular size of the 
Moon varies in a regular way, implying that the Moon's distance varies in a 
regular way \citep{Kri10, Kri16}. The implied eccentricity of the Moon's 
orbit was found to be $\approx$0.04.  The true eccentricity of the Moon's 
orbit is 0.055, but its orbit is anything but a simple ellipse, owing to the 
combined gravitational forces of the Sun and Earth.

Since the time of Hipparchus (ca. 150 BC) it has been known that the minimum 
Earth-Sun distance occurs each year shortly after the winter 
solstice.\footnote[2] {Hipparchus determined that the maximum Earth-Sun 
distance, the ``solar apogee'' in a geocentric model, occurs when the Sun's 
ecliptic longitude is 5.5 degrees east of the boundary between Taurus and 
Gemini \citep[][on p. 211]{hipp}.  This means the ecliptic longitude of the 
Sun at the time of the mininum Earth-Sun distance is roughly 90 + 5.5 + 180 = 
275.5 deg. Accounting for the observed advance of the perihelion of Earth's 
orbit of 11.45 arc \arcsec/yr \citep{fitz}, one finds that the Earth's 
perihelion occurs nowadays in the first few days of January. For a table of 
the Earth's perihelion and aphelion from 2001 to 2100 see: 
http://www.astropixels.com/ephemeris/perap2001.html.} Using very simple 
observations, \citet{Lah12} derived a value of the Earth's orbital 
eccentricity of 0.017 $\pm$ 0.001, which compares extremely well with the 
official modern value of 0.0167.  This was accomplished by determining the 
variation of the equation of time (difference of apparent solar time and mean 
solar time) over the course of the year using observations of the length of 
the shadow of a {\em gnomon}.  It was also necessary to know the obliquity of 
the ecliptic, which is directly obtained from such observations on the first 
day of summer and the first day of winter \citep{Kri_etal12}. The point here 
is that it was certainly known in ancient times and at the time of Copernicus 
that orbiting bodies are not equidistant from the bodies they orbit.

In 1609 Johannes Kepler published the original versions of his first two 
laws of planetary motion: 1) the orbit of a planet is an ellipse, with 
the Sun at one focus; and 2) what we now call the law of areas, that the 
radius vector of a planet sweeps out equal areas in equal times.  The 
Second Law can be stated as follows: 

\begin{equation}
r^2 \rm{d}\theta \; = \; h \; , 
\end{equation}

\noindent
where $r$ is the distance between a planet and the Sun, d$\theta$ is an angular 
increment in radians, and $h$ is a constant unique to each planet.

Newton's breakthroughs in mathematics and mechanics led to the 
realization that Kepler's First Law needed correction. The very center 
of the Sun is not at the focus of a planetary orbit. A planet orbits the 
{\em center of mass} of the planet-Sun system, and the Sun orbits that 
center of mass too \citep[][chapter 2]{Car_Ost07}. This idea, of course, 
has led to the discovery of many extra-solar planets via the radial 
velocity method.

In the autumn of 2015 Mars was nicely situated in the constellation Leo before 
sunrise. We began a sequence of observations of Mars using a simple cross staff 
(Fig. \ref{f1}).\footnote[3]{A pattern for making the cross staff can be obtained 
from this link: 
https://sites.google.com/a/uw.edu/introductory-astronomy-clearinghouse/labs-exercises/measuring-angular-sizes-and-distances. 
The reader should note that when printed out the scale may look like inches, but 
is, in fact, somewhat different.}

Say the full width of the cross staff is $d$, and suppose at a linear distance 
$D$ down the ruler the angular separation of two celestial objects exactly matches 
the width of the cross staff.  Then the angular separation of the two objects will be 

\begin{equation}
\theta \; = \; 2 ~\rm{tan}^{-1} \left(\frac{d}{2D}\right) \; .
\end{equation}

If an observer can measure the angular separation of a planet and two stars 
of known celestial coordinates, there are two possible solutions for the 
position of the planet, one on each side of the great circle arc joining the 
two stars.  If the planet is close to being on the great circle arc between 
the two stars, perhaps no solution results, given errors of measurement.  If 
the positions of a planet and the two stars form a spherical triangle with 
reasonably equal sides, this is ideal, and the planetary position can be 
determined reasonably accurately.  We found that if only two reference stars 
are used, and the observations are not carefully made, the systematic errors 
of the two angular distances can conspire to give a planetary position that 
is in error by more than one degree.  We found it advisable to use three to 
five reference stars. We assume a system of accurate stellar coordinates of 
bright stars along the zodiac.  We adopt the J2000.0 coordinates of such 
stars from the SIMBAD 
database.\footnote[4]{http://simbad.u-strasbg.fr/simbad/}

In Fig. \ref{nov21} we show a nearly ideal set of reference stars
distributed around the location of Mars on 21 November 2018 UT. In our
experience the most accurate angular separations are obtained when the 
reference stars are brighter than fourth magnitude and the angular 
separation, as measured with our cross staff, is between 9 and 15 degrees.   

Understanding Johannes Kepler's efforts to discover the elliptical nature of 
Mars' orbit requires serious effort.  A good place to start is an article by 
\citet{Ging89}.  Kepler endeavored to determine the true shape of a planetary 
orbit. Is it an offset circle, a combination of a larger circle and one epicycle 
(as fit by Copernicus), an ovoid, or an ellipse? To make a long story shorter, 
using data obtained by Tycho Brahe and his assistants, Kepler found systematic 
differences amounting to 8\arcmin ~between the measured ecliptic longitudes of 
Mars and his favored model (an ovoid). These 8\arcmin 
~differences occurred when Mars was located in the octants of its orbit (45 
degrees either side of the Sun, or 135 degrees either side). Since Tycho's 
data were demonstably accurate to $\pm$2\arcmin ~or better, Kepler decided 
that there was a problem with the {\em model}.  This led him to conclude that the 
true orbital shape was an ellipse.

We wondered if it is possible to demonstrate from simple naked eye observations 
that the orbit of Mars is indeed an ellipse.  Or, requiring less rigor, are the 
positions of Mars consistent with an elliptical orbit? If so, can we derive the 
eccentricity?  Here we present results based on two seasons of observing spanning 
410 days and 391 days, respectively.

A full blown orbital determination for the planet Mars is beyond the
scope of the present paper.  That would involve 
simultaneously solving for all six orbital elements.  We only endeavor 
to show that a dataset obtained with simple equipment can be fit with 
an ellipse of eccentricity $\approx$ 0.093.  Other values of the 
eccentricity can be shown to give ecliptic longitudes that differ from 
the observational data by a few degrees, far larger systematic 
differences than the internal random errors of the observations.

\section {Data Acquisition}

In Table \ref{data} we give various data relating to Mars.  For each 
Julian Date we give the ``true'' right ascension ($\alpha$) and declination 
($\delta$) of the planet, obtained using an algorithm of 
\citet{vanF_Pul79}.  These coordinates are accurate to $\pm$1\arcmin.
Note that these coordinates will correspond to the equinox of date.  
To convert these coordinates to ecliptic latitude ($\beta$) and
longitude ($\lambda$) we need the following formulas from spherical
trigonometry \citep[][p. 40]{Smart77}:

\begin{equation}
\rm{sin}(\beta) \; = \; \rm{sin}(\delta) ~\rm{cos}(\epsilon) - 
\rm{cos}(\delta) ~\rm{sin}(\alpha) ~\rm{sin}(\epsilon) \; ;
\end{equation}

\begin{equation}
\rm{sin}(\lambda) \; = \; \frac{\rm{cos}(\delta) ~\rm{sin}(\alpha) ~\rm{cos}(\epsilon) + 
\rm{sin}(\delta) ~\rm{sin}(\epsilon)}{\rm{cos}(\beta)} \; ;
\end{equation}

\begin{equation}
\rm{cos}(\lambda) \; = \; \frac{\rm{cos}(\delta) ~\rm{cos}(\alpha)}{\rm{cos}(\beta)} \; .
\end{equation}

\noindent
$\epsilon$ is the obliquity of the ecliptic, 23\arcdeg ~26\arcmin ~21\arcsec.406
for the year 2000.  Using the {\sc atan2} function in FORTRAN or Python 
with arguments sin($\lambda$) and cos($\lambda$), we obtain the ecliptic 
longitude in the correct quadrant.

Table \ref{data} also gives the observed right ascension, declination, 
ecliptic longitude, and ecliptic latitude of Mars, derived from the cross 
staff measurements, along with the number of reference stars used and the 
value for each date of the Sun's ecliptic longitude. The values of the 
Sun's longitude were calculated using the second method of \citet[][p. 
80]{Meeus88}. On occasion we desired 
one more sufficiently bright reference star and used the {\em derived} 
position of Saturn or Jupiter as the extra reference ``star''.

Consider two celestial objects with equatorial coordinates ($\alpha _1$, $\delta _1$) and
($\alpha _2$, $\delta _2$). The angular separation ($\theta$) between two objects is:

\begin{equation}
\rm{cos} (\theta) \; = \; \rm{sin}(\delta _1) ~\rm{sin}(\delta _2) +
\rm{cos}(\delta _1) ~\rm{cos}(\delta _2) ~\rm{cos}(\alpha _1 - \alpha _2) \; .
\end{equation}

Next consider a spherical quadrilateral that is bounded by starting and ending 
right ascensions, and starting and ending declinations.  The quadrilateral is 
divided into a grid, given a nominal increment in each coordinate of 0.01 deg.  
We used a computer program of our devising that uses the coordinates of two 
reference stars and the measured angular distance of a planet from each of these 
stars to determine the coordinates of the planet.  This is done by brute force,
determining which positions in the quadrilateral match the angular distances
to the two stars, within some settable error.  If the coordinates of the 
stars are J2000.0 coordinates, then the derived right ascension and declination 
of the planet are also J2000.0 coordinates.

The derived ecliptic coordinates of Mars for 2015/2016 are shown in Figure 
\ref{mars_ecl_2016}. The solid line in the plot shows the locus of ``true'' 
positions from \citet{vanF_Pul79}.  The most obvious thing to note is that 
there is a variation of the ecliptic latitude of Mars.  This means that its 
orbit is inclined to the orbit of the Earth.  At different oppostions, Mars 
shows different retrograde patterns on the sky. In this 
paper we will only be analyzing the ecliptic longitudes of Mars vs. time.

Given that most positions of Mars listed in Table \ref{data} were derived 
from angular separations with respect to three to five reference stars, 
almost all out nightly mean derived right ascensions and declinations have 
easy-to-calculate internal random errors. These are sometimes as small as 
$\pm$0.01 deg (which we do not really believe). On one occasion (JD 
2457399.9840) the internal random error for right ascension was $\pm$0.23 
deg and the internal random error of declination was $\pm$0.47 deg.  Typical 
internal random errors for right ascension and declination are $\sigma 
_{\alpha} \approx \sigma _{\delta} \approx \pm~0.10$ deg.

The most objective measure of the accuracy of our data would be the scatter of 
the nightly mean positions with respect to some to-be-determined model.  
However, since we have ``true'' positions of Mars from \citet{vanF_Pul79} we 
can also make a direct estimate of the accuracy of our observations.  The 
easiest way to do this is to precess the ecliptic longitudes from column 4 in 
Table \ref{data} to equinox J2000 by subtracting 50.25 arc seconds per year 
times the number of years from JD 2,457,543.5 (0.0 January 2000) to the date 
of observation.  The ecliptic latitudes require no precession correction.

For the 2015/2016 season the standard deviations of the distributions of 
differences are: $\sigma _{\lambda} = \pm$0.119 deg (7.1\arcmin) for ecliptic 
longitude and $\sigma _{\beta} = \pm$0.178 deg (10.7\arcmin) for ecliptic 
latitude.  The square root of the sum of squares of those errors is $\sigma 
_{tot} = \pm$0.214 deg, or 12.8\arcmin, which we may consider the minimum 
value of the effective accuracy of our cross staff measurements. For the 
2017/2018 season we find $\sigma _{\lambda} = \pm$0.263 deg (15.8\arcmin) and 
$\sigma _{\beta} = \pm$0.201 deg (12.1\arcmin).  Combining the two seasons' 
data we find $\sigma _{\lambda} = \pm$0.185 deg (11.1\arcmin) and $\sigma 
_{\beta} = \pm$0.188 deg (11.3\arcmin).  We are not sure why the second 
season's data are, at face value, less accurate than the data of the first 
season, but the distribution and usefulness of sufficiently bright reference 
stars is not the same for all zodiacal constellations.

\section{Fitting the Data}  

In Book 5, Chapter 19, of {\em On the Revolutions of the Heavenly Spheres} 
\citet{Cop1543} derives the perigee, mean, and apogee distances of Mars.  He 
obtains the values 1.374, 1.520, and 1.649 AU, respectively. Thus, 
Copernicus knew the amount by which Mars' distance from the Sun varies, and 
his mean distance is very close to the modern value of the semi-major axis 
size of Mars' orbit (1.52366 AU).  He used the combination of one large 
circle and one smaller circle, akin to a deferent and an epicycle.

Let us consider the elliptical orbit of Mars.  The equation of an
ellipse is:

\begin{equation}
r \; = \; \frac{a(1 \; - \; e)}{1 \; + \; e \; \rm{cos} \; \theta} \; ,
\end{equation}

\noindent
where r is the Mars-Sun distance, $a$ is the semi-major axis of
the ellipse, $e$ is the eccentricity, and angle $\theta$ = 0 when Mars is 
at perihelion.

The velocity along the orbit \citep[][Eq. 2.36]{Car_Ost07} is

\begin{equation}
v^2 \; = \; G (M_{\odot} + M_{Mars}) \left(\frac{2}{r} \; - \; \frac{1}{a} \right) \; .
\end{equation}

\noindent
Since the mass of Mars is $\approx$ 3.23 $\times$ 10$^{-7}$ M$_{\odot}$
\citep[][p. 295]{Tho_etal00}, for our purposes here we shall ignore it.

We wish to calculate the position of Mars at increments of one Earth day 
starting at the moment of its perihelion.  At perihelion r = r$_{min}$ = $a 
(1 - e)$. Using the known semi-major axis size and eccentricity of Mars' 
orbit (or a range of assusmed values), we can calculate the maximum velocity 
at perihelion with Equation 8. On perihelion day traveling at velocity 
$v_{max}$ Mars moves 0.6349 degrees along its orbit as viewed from the Sun. 
This allows us to calculate the constant $h$ for Mars using Equation 1. 
Then, by alternating use of Equations 7 and 1 we can calculate $r$ and 
$\theta$ for Mars each day along its orbit.  The X-Y coordinates are 
obtained simply: $X = r$ cos($\theta$) and $Y = r$ sin($\theta$).

Given the small eccentricity of the Earth's orbit (0.0167), let us begin 
with the simplest possible model by assuming a circular orbit for the Earth, 
which of course implies a constant velocity. On average, the Earth moves 
360.0/365.2422 = 0.985647 degrees per day with respect to the Sun.  This 
gives us another set of X-Y coordinates in the same coordinate system, with 
the Sun at the origin and the +X-axis in the direction of Mars' perihelion.

Consider Fig. \ref{oppos}.  According to the Solar Systems Dynamic Group, 
Horizons On-Line Ephemeris System at Jet Propulsion Laboratory,\footnote[5]{
https://ssd.jpl.nasa.gov/horizons.cgi} one perihelion 
of Mars occurred on Julian Date 2,457,003.8524 (12 December 2014, at 08:27:29 
UT).  The next perihelion occurred on Julian Date 2,457,691.0507 (29 October 
2016, at 13:13:04 UT).  For the moment let us take the perihelion dates as 
given.  They are not directly observable, but the date of opposition of Mars 
essentially is.  Mars' ecliptic longitude differs from that of the Sun by 180 
deg near the mid-time of its retrograde motion. From our data we find 
opposition to have occurred at JD 2,457,730.36 $\pm$ 0.62 (21 May 2016 at 21 
hours UT). According to the {\em Astronomical Almanac for the Year 2016}, 
opposition occurred on May 22nd at 11 hours UT, or JD 2,457,530.96.  The 
agreement is within one standard deviation.

Mars' 2016 opposition occurred 526.9 days after the perihelion of 12 December 
2014.  Call it 527 days.  The X-Y coordinates of the day-by-day position of Mars 
in our coordinate system give $\theta$ = 265.545 deg on the day of opposition.  
Since the Earth moves 0.985647 deg per day along its orbit, the 269th pair of 
Earth coordinates gives an angle $\theta$ most closely matching that of Mars 
(265.139 vs. 265.545 deg, in fact).  Given the index $i$ of the Earth's 
coordinates, the corresponding index of Mars' coordinates for the same day is 
equal to $i$ + (527 $-$ 269).

We wrote a simple Python program that calculates the X-Y coordinates of Mars 
for each day starting at perihelion, and the X-Y coordinates of the Earth.  
With the appropriate offset of the indices of the two sets of coordinates we 
can obtain the direction toward Mars {\em from the Earth} for any given 
date. This generates a locus of ``some angle'' vs. time in days.  We then 
used a FORTRAN program originally written for fitting templates to supernova 
light curve photometry to adjust that locus to the dates and ecliptic 
longitudes of our Mars observations.  This produces a goodness of fit 
parameter equal to the sum of squares of differences between the template 
and the data (in other words, like $\chi^2$ minimization, but with equal 
weights for all the points).  The square root of the goodness of fit 
parameter divided by the number of data points minus two gives the RMS 
residual.

We tried values of the eccentricity of Mars' orbit ranging from 0.053 to 
0.133. As shown in the bottom panel of Fig. \ref{mars_2016}, for e = 0.053 
the observed residuals are +1.0 deg at the start, $-$1.3 deg at JD 
2,457,624, and +3.4 deg at the end. For e = 0.133 the observed residuals 
are $-$1.9 deg at the start, but +1.9 deg at JD 2,457,600, and $-$1.3 deg 
at the end.  A fit with e = 0.0934 (the modern accepted value) is much 
better. The 2015/2016 dataset and the use of a circular orbit for the Earth give a 
most likely value of eccentricity of 0.0930 $\pm$ 0.012. Thus, our simple 
model allows us to show that data much less accurate than Tycho's are 
consistent with an elliptical orbit for Mars, and one having an 
eccentricity equal to the modern accepted value, within the errors.

For the best fit the RMS residual is $\pm$0.297 deg (17.8\arcmin). There is 
a trend to the residuals. The are primarily negative at the start, positive 
at the end, with the final residual equal to +1.0 deg $-$ a 3-$\sigma$ 
outlier.  What improvements can result from a model that adopts an 
elliptical shape for the Earth's orbit, with e = 0.0167?  To do this we 
need to know the day of the year when opposition occurs (which we have 
already determined) and the day of the year of the Earth's perihelion, 
which was 3 January 2016 to the nearest day. We then rotate the grid of X-Y 
coordinates of the position of the Earth so that the angle from the Sun 
towards the Earth matches the angle of the Sun towards Mars on opposition 
day, since all three must line up on that day. The differences $X_{Mars} - 
X_{Earth}$ and $Y_{Mars} - Y_{Earth}$ are passed to the ATAN2 function as 
the two arguments. We generate a locus lasting more than 410 days, as that 
is the extent of our dataset.  The resultant locus is then shifted using 
the template fitting program to give the best fit of the model to the data.  
Using the modern accepted values of the eccentricity of Mars, time of 
perihelion in 2014, and semi-major axis size for Mars, this gives an RMS 
residual of $\pm$ 7.5\arcmin.  See Fig. \ref{mars_ecc_2016}.

In Fig. \ref{mars_ecl_2018} we show the observed positions of Mars
from the 2017/2018 season, along with the values derived from the
algorithm of \citet{vanF_Pul79}.  Clearly, the retrograde loop is quite
different than that of the 2015/2016 season (Fig. \ref{mars_ecl_2016}).
We note that in 2015/2016 the observed duration of retrograde motion of 
Mars was 73.34 days.  Mars backed up 16.41 $\pm$ 0.26 deg.  In 2017/2018 
retrograde motion lasted only 61.92 days, and Mars backed up 10.65 $\pm$ 0.26 deg.
Knowing this, we could predict that a locus that fits the ecliptic longitudes
of 2015/2016 cannot just be shifted in time and angle to fit the ecliptic
longitudes of 2017/2018.

For the 2017/2018 season the opposition of Mars was observed to occur at JD 
2,458,325.14 $\pm$ 0.96 (25.64 July 2018 UT).  According to the {\em 
Astronomical Almanac for the Year 2018}, opposition occurred on 27.21 July 2018 
UT, or JD 2,458,326.71. The agreement is within 1.64-$\sigma$.

Fig. \ref{mars_ecc_2018} shows two fits for the 2017/2018 data.  The simpler
model, using a uniform circular orbit for the Earth, clearly does not work.
The residuals vary from $-$9.0 deg at the start to +7.7 deg after opposition.
The RMS residual is $\pm$6.2 deg.  For our second season of data we must
use elliptical orbits for both Mars and the Earth.  This fit (the solid line
in Fig. \ref{mars_ecc_2018}) gives an RMS residual for the ecliptic longitudes
of $\pm$0.202 deg (12.1\arcmin).  

Why does the simpler model not work for the 2017/2018 data?  Consider Fig. 
\ref{oppos}. In 2018 the aphelion of the Earth occurred on July 6th. 
Opposition of Mars occurred on July 25 (observed) or July 27 (true), just 19 
to 21 days later.  So the Earth was moving almost as slowly as it does along 
its orbit.  Mars was at perihelion on 16 September 2018 at 07:54 
UT.\footnote[6]{https://in-the-sky.org/news.php?id=20180916\_12\_100} At 
opposition it was 33 deg short of being at perihelion, meaning that Mars was 
moving just about as fast along its orbit as possible.  During 2015/2016 Mars 
was near the minor axis of its orbit at the time of opposition, moving at a 
velocity close to its mean velocity along its orbit. A circular orbit for 
the Earth gave a reasonably good fit, but the trend of the residuals over
a timespan greater than one year motivated us to improve the fitting by
also using an elliptical orbit for the Earth.

\section{Conclusions}

Our value of the orbital eccentricity of Mars (0.093 $\pm$ 0.012), derived 
using the simpler fit (a circular Earth orbit), compares well with the 
accepted modern value (0.0934).  While our data are not accurate enough to 
{\em prove} that Mars' orbit is an ellipse, if we fit the data with an 
ellipse, we can show that the eccentricity must be near 0.09.  Thus, a dataset 
based on naked eye observations accurate to 0.2 or 0.3 deg in ecliptic 
longitude can be shown to be in agreement with Kepler's First Law.

During 2018 the opposition of Mars occurred not long after Earth's aphelion 
and less than two months before Mars' perihelion.  These conditions obliged 
us to fit the ecliptic longitudes of Mars of our second season using an 
elliptical orbit for Mars and an elliptical orbit for Earth. Refitting the 
2015/2016 ecliptic longitudes of Mars using an ellipse for Mars' orbit and 
an ellipse for the Earth gives an RMS uncertainty of $\pm$7.5\arcmin. We 
regard this as a suprisingly accurate result given the primitive nature of 
our simple cross staff.

\acknowledgments

We made use of the SIMBAD database, operated at CDS, Strasbourg, France.

\clearpage

\begin{deluxetable}{lrrrrrrrccc}
\tablecolumns{9}
\tablewidth{0pc}
\tabletypesize{\scriptsize}
\tablecaption{Mars Data\label{data}}
\tablehead{ \colhead{JD\tablenotemark{a}} &
\colhead{RA$_{true}$\tablenotemark{b}} & 
\colhead{DEC$_{true}$\tablenotemark{b}} & 
\colhead{$\lambda$\tablenotemark{c}} & 
\colhead{$\beta$\tablenotemark{d}} & 
\colhead{RA$_{obs}$} & 
\colhead{DEC$_{obs}$} & 
\colhead{$\lambda _{obs}$} &
\colhead{$\beta _{obs}$} &
\colhead{N\tablenotemark{e}} & 
\colhead{$\lambda _{\odot}$} }
\startdata
%
%

7338.9674 & 180.3206 & \phn1.4016 & 179.737 & \phn1.413 & 180.31  & \phn1.68 & 179.62 & \phn1.67 &  2  & 229.748 \\
7344.9931 & 183.6448 &  $-$0.0205 & 183.353 & \phn1.430 & 183.35  & \phn0.10 & 183.03 & \phn1.42 &  4  & 235.817 \\
7346.9708 & 184.7339 &  $-$0.4856 & 184.538 & \phn1.435 & 184.58  &  $-$0.26 & 184.31 & \phn1.59 &  2  & 237.812 \\
7359.9819 & 191.8332 &  $-$3.4881 & 192.243 & \phn1.467 & 191.65  &  $-$3.57 & 192.11 & \phn1.32 &  4  & 250.977 \\
7364.9833 & 194.5444 &  $-$4.6133 & 195.173 & \phn1.477 & 194.12  &  $-$4.34 & 194.68 & \phn1.57 &  3  & 256.053 \\
7370.9813 & 197.7750 &  $-$5.9320 & 198.651 & \phn1.488 & 197.45  &  $-$5.95 & 198.36 & \phn1.35 &  4  & 262.149 \\
7373.9674 & 199.3779 &  $-$6.5758 & 200.370 & \phn1.492 & 199.19  &  $-$6.50 & 200.17 & \phn1.49 &  4  & 265.187 \\
7380.9542 & 203.1074 &  $-$8.0433 & 204.351 & \phn1.501 & 203.01  &  $-$8.09 & 204.28 & \phn1.42 &  4  & 272.301 \\
7387.9889 & 206.8359 &  $-$9.4620 & 208.301 & \phn1.507 & 206.51  &  $-$9.44 & 207.99 & \phn1.41 &  4  & 279.469 \\
7391.9813 & 208.9367 & $-$10.2376 & 210.512 & \phn1.509 & 208.73  & $-$10.12 & 210.28 & \phn1.55 &  4  & 283.538 \\
\\
7399.9840 & 213.1084 & $-$11.7237 & 214.870 & \phn1.508 & 212.89  & $-$11.56 & 214.61 & \phn1.59 &  4  & 291.694 \\
7406.9861 & 216.7061 & $-$12.9424 & 218.591 & \phn1.503 & 216.45  & $-$12.77 & 218.30 & \phn1.59 &  4  & 298.986 \\
7410.9840 & 218.7332 & $-$13.6022 & 220.672 & \phn1.497 & 218.60  & $-$13.08 & 220.39 & \phn1.95 &  5  & 302.894 \\
7415.9944 & 221.2445 & $-$14.3918 & 223.235 & \phn1.488 & 220.94  & $-$14.19 & 222.89 & \phn1.59 &  4  & 307.990 \\
7425.9861 & 226.1101 & $-$15.8320 & 228.149 & \phn1.460 & 225.91  & $-$15.74 & 227.94 & \phn1.50 &  5  & 318.128 \\
7429.9715 & 227.9932 & $-$16.3574 & 230.034 & \phn1.445 & 227.85  & $-$16.34 & 229.90 & \phn1.43 &  5  & 322.164 \\
7435.9937 & 230.7561 & $-$17.0962 & 232.782 & \phn1.417 & 230.26  & $-$17.05 & 232.31 & \phn1.34 &  5  & 328.250 \\
7443.9931 & 234.2464 & $-$17.9769 & 236.225 & \phn1.369 & 233.84  & $-$17.98 & 235.85 & \phn1.28 &  5  & 336.312 \\
7449.9799 & 236.6977 & $-$18.5632 & 238.626 & \phn1.322 & 236.20  & $-$18.77 & 238.21 & \phn1.02 &  5  & 342.326 \\
7451.9677 & 237.4743 & $-$18.7441 & 239.383 & \phn1.305 & 236.99  & $-$19.08 & 239.01 & \phn0.88 &  5  & 344.318 \\
\\
7467.9639 & 242.9168 & $-$19.9740 & 244.661 & \phn1.114 & 242.59  & $-$20.01 & 244.37 & \phn1.02 &  5\tablenotemark{f} & \phn\phn0.280 \\
7472.9521 & 244.2518 & $-$20.2824 & 245.949 & \phn1.033 & 243.94  & $-$20.39 & 245.68 & \phn0.88 &  5\tablenotemark{f} & \phn\phn5.228 \\
7480.9486 & 245.9309 & $-$20.7106 & 247.572 & \phn0.874 & 245.74  & $-$20.76 & 247.40 & \phn0.80 &  5\tablenotemark{f} & \phn13.132 \\
7499.9549 & 247.1279 & $-$21.4551 & 248.792 & \phn0.316 & 246.72  & $-$21.55 & 248.43 & \phn0.16 &  4  & \phn31.773 \\
7512.9493 & 245.2332 & $-$21.7269 & 247.097 &  $-$0.235 & 244.72  & $-$21.72 & 246.63 &  $-$0.31 &  5  & \phn44.407 \\
7515.9406 & 244.4834 & $-$21.7551 & 246.416 &  $-$0.381 & 244.07  & $-$21.62 & 246.01 &  $-$0.31 &  5  & \phn47.304 \\
7538.6354 & 236.5451 & $-$21.4892 & 239.126 &  $-$1.564 & 236.29  & $-$21.59 & 238.92 &  $-$1.72 &  5\tablenotemark{f} & \phn69.160 \\
7547.6240 & 233.4845 & $-$21.2476 & 236.295 &  $-$1.979 & 233.11  & $-$21.32 & 235.97 &  $-$2.13 &  3  & \phn77.768 \\
7555.6153 & 231.4404 & $-$21.0834 & 234.404 &  $-$2.282 & 231.00  & $-$21.20 & 234.03 &  $-$2.50 &  5  & \phn85.406 \\
7564.6441 & 230.1675 & $-$21.0421 & 233.243 &  $-$2.541 & 229.98  & $-$20.87 & 233.03 &  $-$2.42 &  5  & \phn94.022 \\
\\
7569.6226 & 229.9735 & $-$21.1049 & 233.084 &  $-$2.648 & 229.67  & $-$21.42 & 232.89 &  $-$3.03 &  5  & \phn98.770 \\
7576.6399 & 230.3070 & $-$21.3033 & 233.436 &  $-$2.760 & 230.12  & $-$21.13 & 233.22 &  $-$2.64 &  5  & 105.460 \\
7582.6149 & 231.1228 & $-$21.5684 & 234.238 &  $-$2.825 & 230.94  & $-$21.37 & 234.02 &  $-$2.68 &  5  & 111.158 \\
7591.6299 & 233.1991 & $-$22.1097 & 236.242 &  $-$2.880 & 232.79  & $-$22.22 & 235.90 &  $-$3.08 &  5  & 119.759 \\
7597.6517 & 235.0988 & $-$22.5434 & 238.054 &  $-$2.891 & 234.89  & $-$22.38 & 237.83 &  $-$2.78 &  4\tablenotemark{f} & 125.512 \\
7602.6056 & 236.9372 & $-$22.9283 & 239.795 &  $-$2.889 & 236.56  & $-$23.13 & 239.50 &  $-$3.16 &  5\tablenotemark{f} & 130.249 \\
7624.6229 & 247.6334 & $-$24.6631 & 249.743 &  $-$2.783 & 247.18  & $-$24.45 & 249.30 &  $-$2.64 &  5  & 151.392 \\
7633.5979 & 252.9637 & $-$25.2414 & 254.615 &  $-$2.707 & 252.55  & $-$25.30 & 254.25 &  $-$2.81 &  5  & 160.063 \\
7639.5740 & 256.7638 & $-$25.5420 & 258.065 &  $-$2.649 & 256.41  & $-$25.57 & 257.75 &  $-$2.71 &  5  & 165.857 \\
7646.5729 & 261.4329 & $-$25.7851 & 262.284 &  $-$2.575 & 261.01  & $-$25.79 & 261.90 &  $-$2.60 &  5  & 172.665 \\
\\
7653.5681 & 266.2981 & $-$25.8906 & 266.667 &  $-$2.494 & 266.05  & $-$25.93 & 266.45 &  $-$2.54 &  5  & 179.495 \\
7660.6153 & 271.3719 & $-$25.8399 & 271.236 &  $-$2.407 & 270.84  & $-$26.06 & 270.76 &  $-$2.62 &  5  & 186.403 \\
7665.5632 & 275.0146 & $-$25.7023 & 274.521 &  $-$2.342 & 274.62  & $-$25.78 & 274.16 &  $-$2.41 &  5  & 191.269 \\
7677.5806 & 275.0688 & $-$24.9923 & 282.737 &  $-$2.173 & 283.87  & $-$24.62 & 282.59 &  $-$1.79 &  4  & 203.147 \\
7689.5389 & 293.2365 & $-$23.7374 & 291.184 &  $-$1.991 & 293.00  & $-$23.67 & 290.98 &  $-$1.89 &  3  & 215.049 \\
7709.5990 & 308.6071 & $-$20.4297 & 305.801 &  $-$1.660 & 308.14  & $-$20.52 & 305.36 &  $-$1.64 &  3  & 235.175 \\
7714.5486 & 312.3536 & $-$19.3998 & 309.471 &  $-$1.575 & 312.04  & $-$19.75 & 309.09 &  $-$1.83 &  5  & 240.170 \\
7721.5361 & 317.5932 & $-$17.8184 & 314.683 &  $-$1.453 & 317.24  & $-$18.03 & 314.39 &  $-$1.59 &  4  & 247.239 \\
7748.5375 & 337.2173 & $-$10.5854 & 335.016 &  $-$0.981 & 337.22  & $-$10.29 & 335.13 &  $-$0.71 &  4  & 274.685 \\
\hline
\\
8085.9910 & 201.8528 &  $-$8.0165 & 203.186 &    1.070  & 201.60  &  $-$7.87 & 202.90 &  \phn1.11 &  3  & 246.441 \\
8127.9736 & 227.2510 & $-$16.7716 & 229.459 &    0.856 & 227.20 & $-$16.81 & 229.42 &  \phn0.81 & 4 & 289.151 \\
8142.0066 & 236.0592 & $-$19.0221 & 238.137 &    0.742 & 235.67 & $-$19.34 & 237.85 &  \phn0.35 & 5 & 303.441 \\
8149.9903 & 241.1237 & $-$20.1117 & 243.031 &    0.665 & 240.79 & $-$20.09 & 242.72 &  \phn0.63 & 4\tablenotemark{g} & 311.555 \\
8189.9872 & 266.6963 & $-$23.3068 & 266.966 &    0.098 & 266.46 & $-$23.08 & 266.74 &  \phn0.32 & 4 & 351.854 \\
8197.9688 & 271.7025 & $-$23.4952 & 271.561 & $-$0.065 & 271.58 & $-$22.90 & 271.46 &  \phn0.53 & 4 & 359.803 \\
8212.9583 & 280.8410 & $-$23.5000 & 279.933 & $-$0.434 & 281.06 & $-$23.45 & 280.14 & $-$0.40 & 4 & \phn14.635 \\
8225.9274 & 288.3395 & $-$23.2071 & 286.811 & $-$0.832 & 288.25 & $-$23.25 & 286.72 & $-$0.86 & 4 & \phn27.366 \\
8235.9472 & 293.7655 & $-$22.8568 & 291.804 & $-$1.198 & 293.80 & $-$22.65 & 291.87 & $-$1.00 & 3\tablenotemark{f} & \phn37.139 \\
8257.7462 & 303.9944 & $-$22.0331 & 301.243 & $-$2.205 & 303.69 & $-$22.25 & 300.92 & $-$2.35 & 4 & \phn58.232 \\
\\
8304.9201 & 312.7415 & $-$23.1844 & 308.798 & $-$5.317 & 312.24 & $-$23.44 & 308.28 & $-$5.44 & 5 & 103.360 \\
8312.9441 & 311.5969 & $-$23.9966 & 307.564 & $-$5.818 & 311.13 & $-$24.05 & 307.14 & $-$5.76 & 5 & 111.011 \\
8332.6215 & 306.5237 & $-$25.8726 & 302.611 & $-$6.474 & 305.93 & $-$26.20 & 302.01 & $-$6.67 & 4 & 129.802 \\
8337.6104 & 305.1761 & $-$26.1587 & 301.359 & $-$6.468 & 304.24 & $-$26.43 & 300.47 & $-$6.54 & 4 & 134.579 \\
8345.6142 & 303.4003 & $-$26.3727 & 299.749 & $-$6.317 & 302.88 & $-$26.55 & 299.25 & $-$6.39 & 4 & 142.257 \\
8361.5858 & 302.2008 & $-$25.9168 & 298.792 & $-$5.639 & 301.54 & $-$26.08 & 298.17 & $-$5.67 & 4 & 157.648 \\
8367.5785 & 302.6835 & $-$25.4859 & 299.311 & $-$5.311 & 302.42 & $-$25.52 & 299.07 & $-$5.29 & 4 & 163.451 \\
8387.5573 & 307.5650 & $-$23.2487 & 304.169 & $-$4.155 & 307.30 & $-$23.41 & 303.89 & $-$4.25 & 5 & 182.927 \\
8399.6042 & 312.3694 & $-$21.3794 & 308.954 & $-$3.486 & 312.25 & $-$20.89 & 308.98 & $-$2.98 & 4 & 194.776 \\
8418.5285 & 321.6447 & $-$17.7297 & 318.389 & $-$2.538 & 321.26 & $-$17.78 & 318.02 & $-$2.47 & 4 & 213.554 \\
\\
8425.5278 & 325.4224 & $-$16.1759 & 322.313 & $-$2.223 & 325.46 & $-$16.25 & 322.32 & $-$2.31 & 4 & 220.549 \\
8443.5750 & 335.6485 & $-$11.7441 & 333.160 & $-$1.499 & 335.40 & $-$11.77 & 332.92 & $-$1.44 & 5 & 238.697 \\
8449.5701 & 339.1479 & $-$10.1586 & 336.938 & $-$1.286 & 339.30 &  $-$9.97 & 337.15 & $-$1.17 & 5 & 244.757 \\
8454.5538 & 342.0836 &  $-$8.8063 & 340.131 & $-$1.119 & 342.02 &  $-$8.61 & 340.15 & $-$0.91 & 5 & 249.805 \\
8463.6052 & 347.4590 &  $-$6.2888 & 346.018 & $-$0.839 & 345.93 &  $-$6.28 & 345.93 & $-$0.79 & 4 & 258.993 \\
8476.5948 & 355.2602 &  $-$2.5803 & 354.627 & $-$0.485 & 355.10 &  $-$2.56 & 354.49 & $-$0.40 & 5 & 272.214 \\
\enddata
\tablecomments{Except for columns 1 and 10, all values are measured in degrees.
A text file containing the data, along with similar data for Venus, Jupiter, and Saturn,
can be obtained via http://people.physics.tamu.edu/public\_html/planets.txt.}
\tablenotetext{a}{Julian Date minus 2,450,000.}
\tablenotetext{b}{True right ascension and declination for equinox of date, 
using algorithm of \citet{vanF_Pul79}.
Accurate to $\pm$1\arcmin.}
\tablenotetext{c}{True ecliptic longitude for equinox of date.  To compare column 4 values to those in column 8,
the column 4 values must be precessed to equinox J2000.0 by subtracting $\approx$0.230\arcdeg (50.25 arc seconds
per year) for the 2015/2016 observing season, and $\approx$0.258\arcdeg for the 2017/2018 observing season.}
\tablenotetext{d}{True ecliptic latitude.  These can be compared directly with the values in column 9, without
any precession.}
\tablenotetext{e}{Number of reference stars used.}
\tablenotetext{f}{One of the reference ``stars'' used was Saturn, based on the position {\em derived} from our observations.}
\tablenotetext{g}{One of the reference ``stars'' used was Jupiter, based on the position {\em derived} from our observations.}
\end{deluxetable}



\clearpage

\figcaption[f1.eps]
{The cross staff.  The cardboard cross piece slides up and down the
yardstick.  Using simple geometry we can use this device to determine
the angular separation of two objects in the sky.  
\label{f1}
}

\figcaption[nov21.eps]
{Position of Mars on 20/21 November 2018 with respect to five bright stars in
Capricornus and Aquarius.  The lengths of the ``spokes'' of the wheel are equal
to the angular separations of Mars and the five stars, as measured with the
cross staff.  Except for $\lambda$ Aqr plus $\beta$ Aqr and $\lambda$ Aqr plus 
$\delta$ Cap, the other seven combinations of two reference stars can be 
used to determine the position of Mars.
\label{nov21}
}

\figcaption[mars_ecl_2016.eps]
{Observed positions of Mars from 12 November 2015 to 26 December 2016 UT. 
The solid line shows the true positions, as derived using the algorithm 
of \citet{vanF_Pul79}.
\label{mars_ecl_2016}
}

\figcaption[oppos.eps]
{The outer locus represents the orbit of Mars.  The X-axis lies along the 
major axis of Mars' ellipse.  The inner locus represents the orbit of the Earth. 
Mars was observed to be at opposition on 22 May 2016 and 25 July 2018.  
Mars' perihelion position is labeled ``p''. 
\label{oppos}
}

\figcaption[mars_2016.eps]
{Observed positions of Mars in 2015/2016 (upper panel). We show three fits to 
the data, using values of eccentricity for Mars' orbit of 0.053, 0.093, and 
0.133. The Earth's orbit is assumed to be circular. The residuals of the three 
fits are shown in the bottom panel and indicate that the fit with eccentricity 
0.093 is clearly the best of the three.
\label{mars_2016}
}

\figcaption[mars_ecc_2016.eps]
{Observed positions of Mars in 2015/2016 and a fit to the data assuming
elliptical orbits for Mars and the Earth.  The RMS residual of the fit
is $\pm$0.125 deg (7.5 arc minutes).
\label{mars_ecc_2016}
}

\figcaption[mars_ecl_2018.eps]
{Observed positions of Mars from 28 November 2017 to 24 December 2018 UT. 
The solid line shows the true positions, as derived using the algorithm 
of \citet{vanF_Pul79}.
\label{mars_ecl_2018}
}

\figcaption[mars_ecc_2018.eps]
{Observed positions of Mars in 2017/2018 and two fits.  The dashed line
was calculated assuming a circular orbit for the Earth.  The  solid line 
was calculated assuming elliptical orbits for Mars and the Earth.  The RMS 
residual of the two ellipse fit is $\pm$0.202 deg (12.1 arc minutes).
\label{mars_ecc_2018}
}

\clearpage

\begin{figure}
\plotone{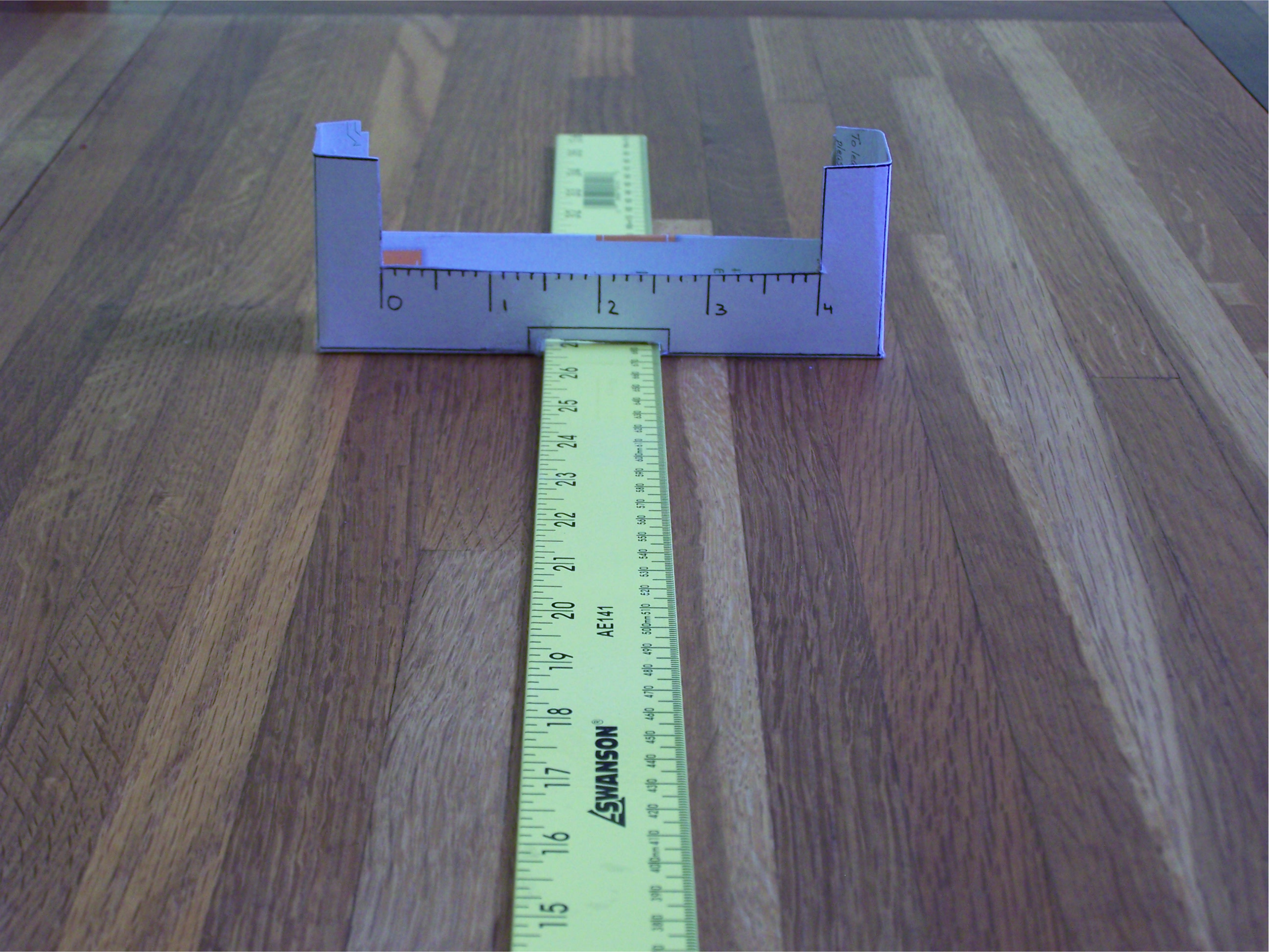} {Krisciunas Fig. \ref{f1}. 
}
\end{figure}

\begin{figure}
\plotone{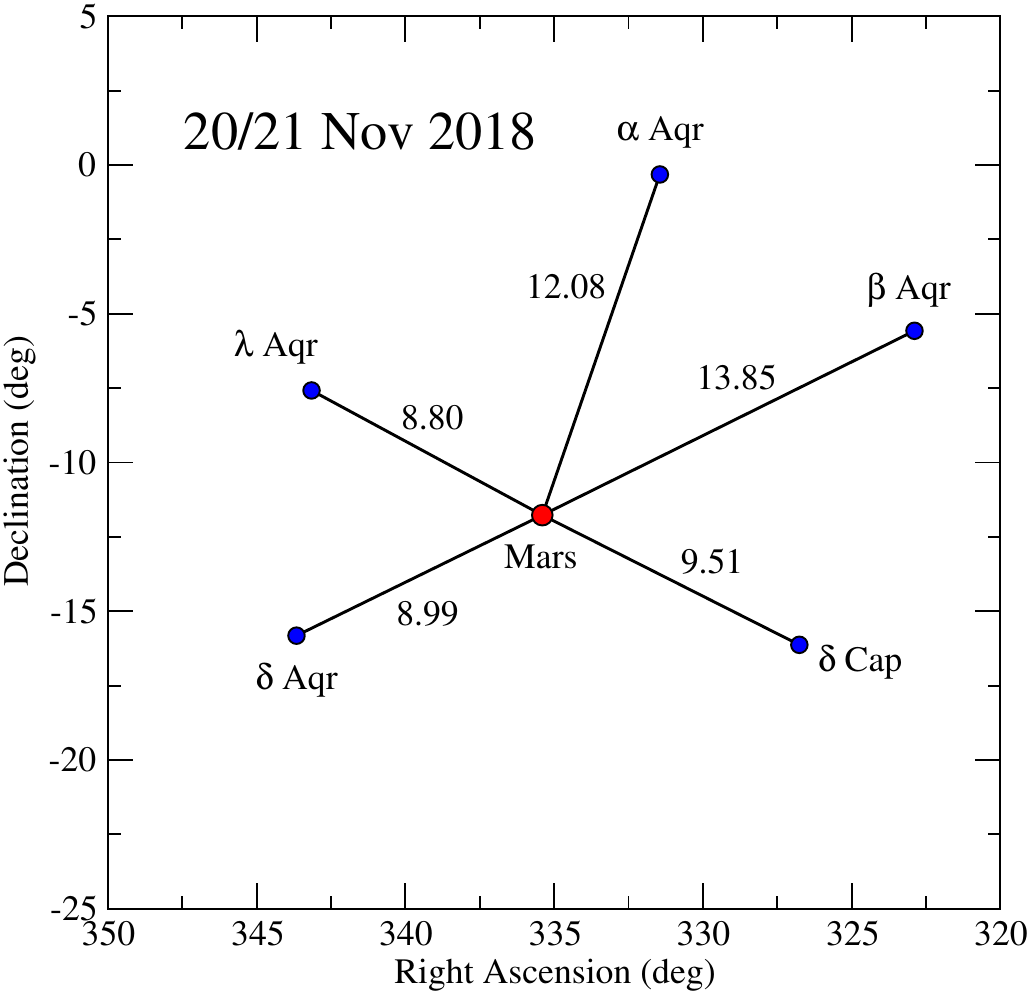} {Krisciunas Fig. \ref{nov21}. 
}
\end{figure}

\begin{figure}
\plotone{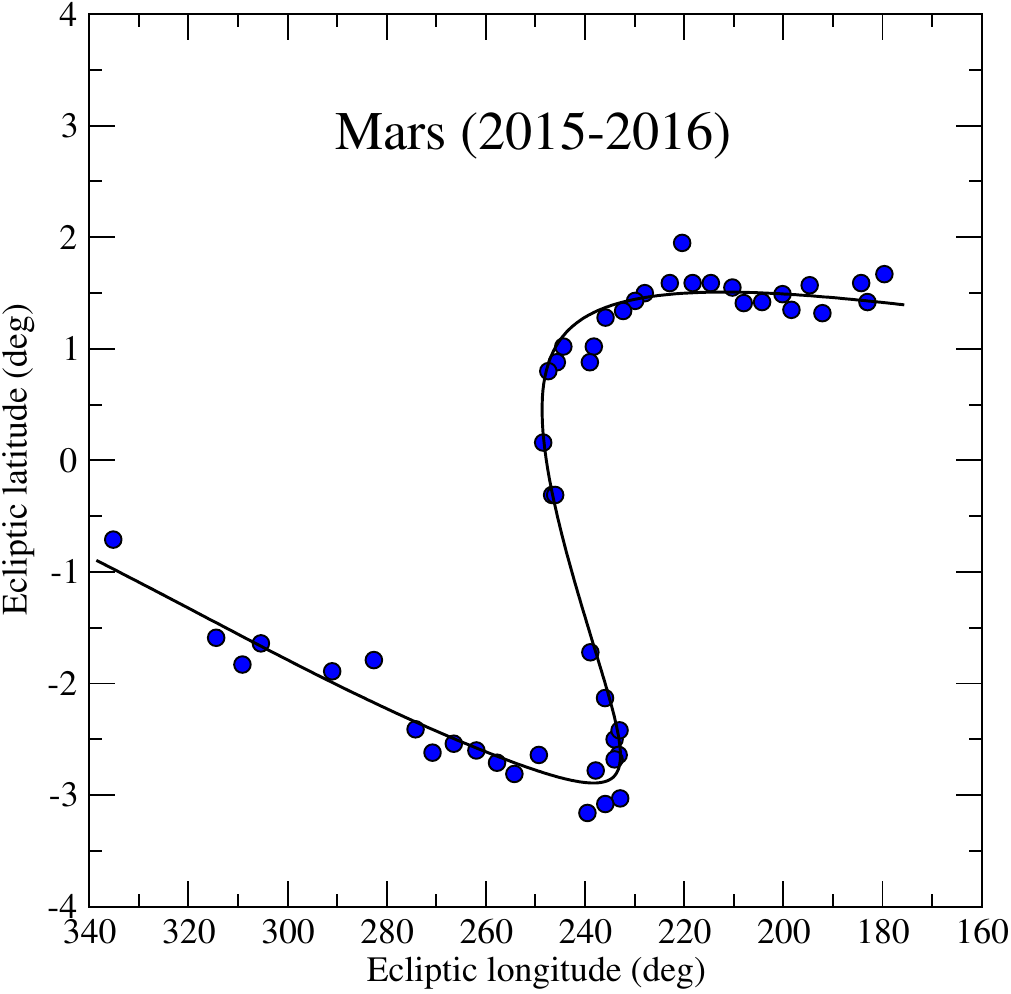} {Krisciunas Fig. \ref{mars_ecl_2016}. 
}
\end{figure}

\begin{figure}
\plotone{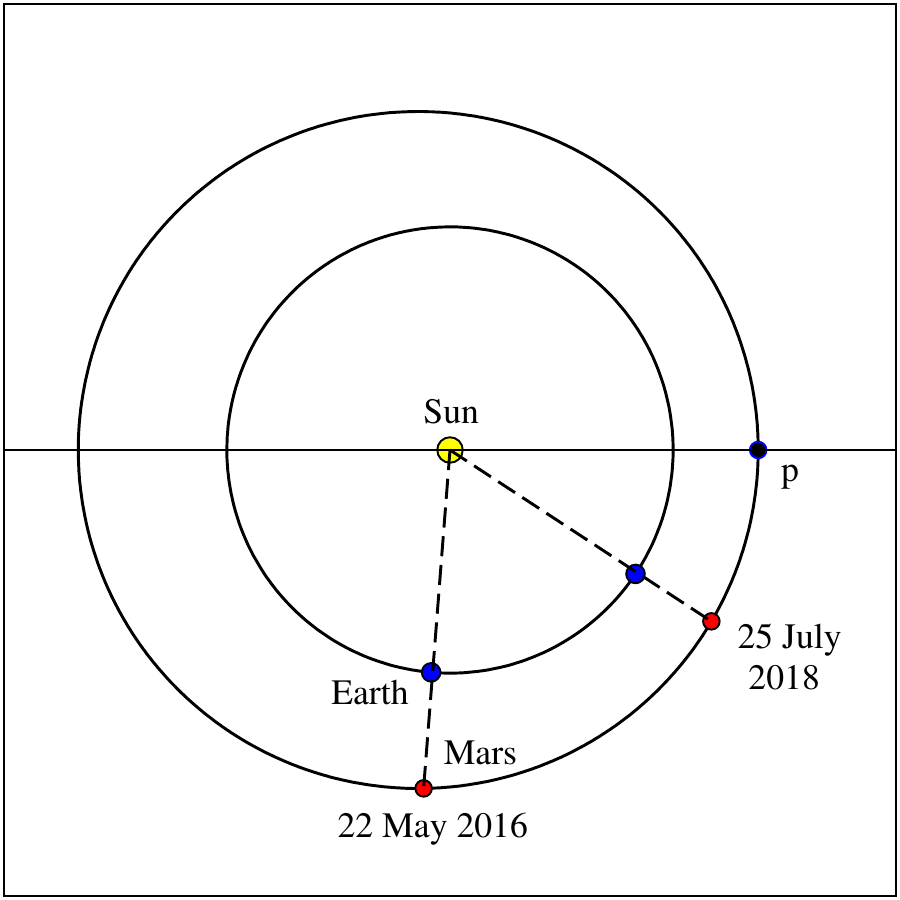} {Krisciunas Fig. \ref{oppos}. 
}
\end{figure}

\begin{figure}
\plotone{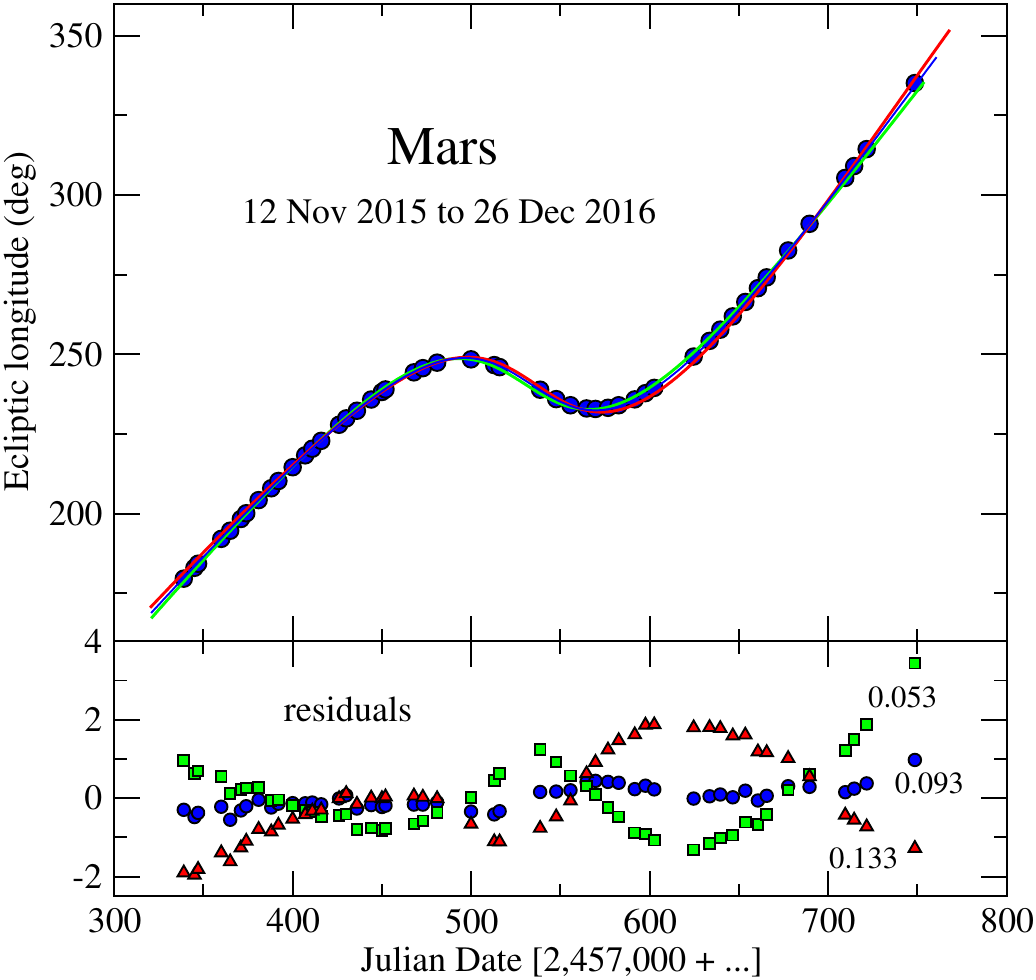} {Krisciunas Fig. \ref{mars_2016}. 
}
\end{figure}

\begin{figure}
\plotone{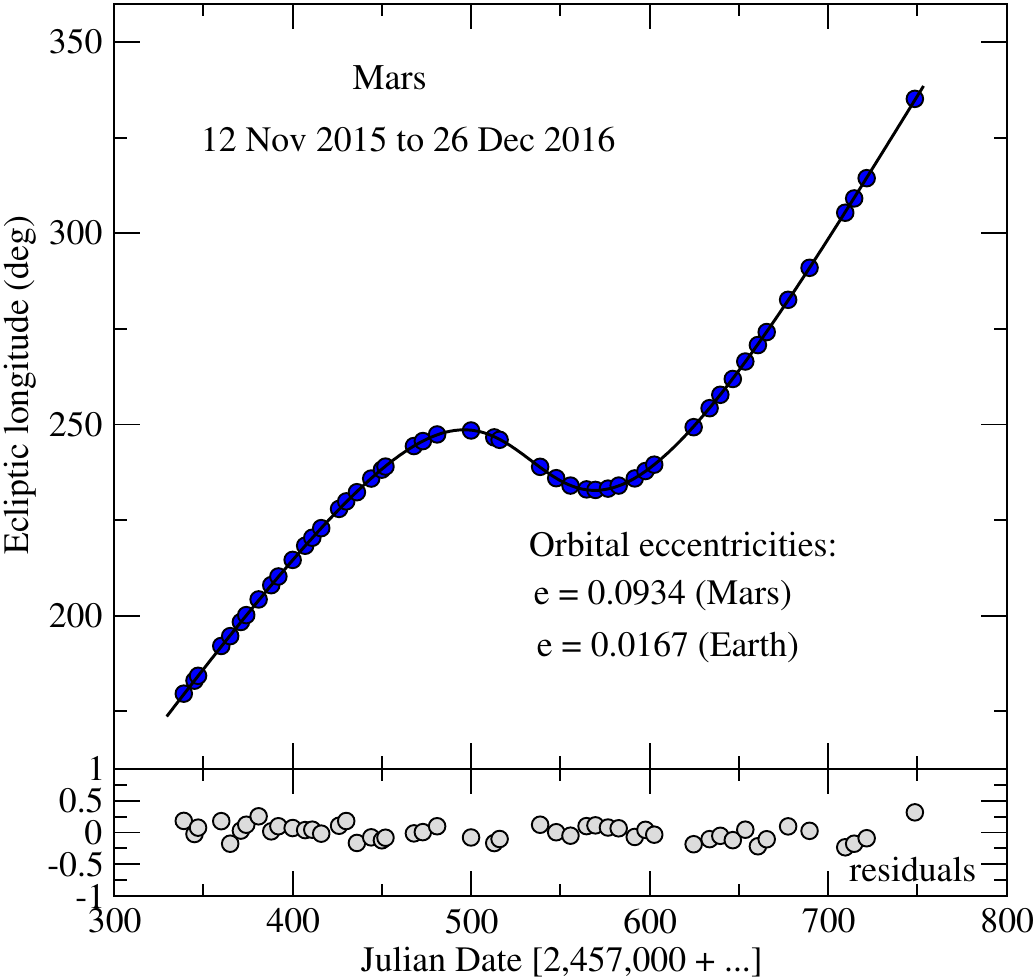} {Krisciunas Fig. \ref{mars_ecc_2016}. 
}
\end{figure}

\begin{figure}
\plotone{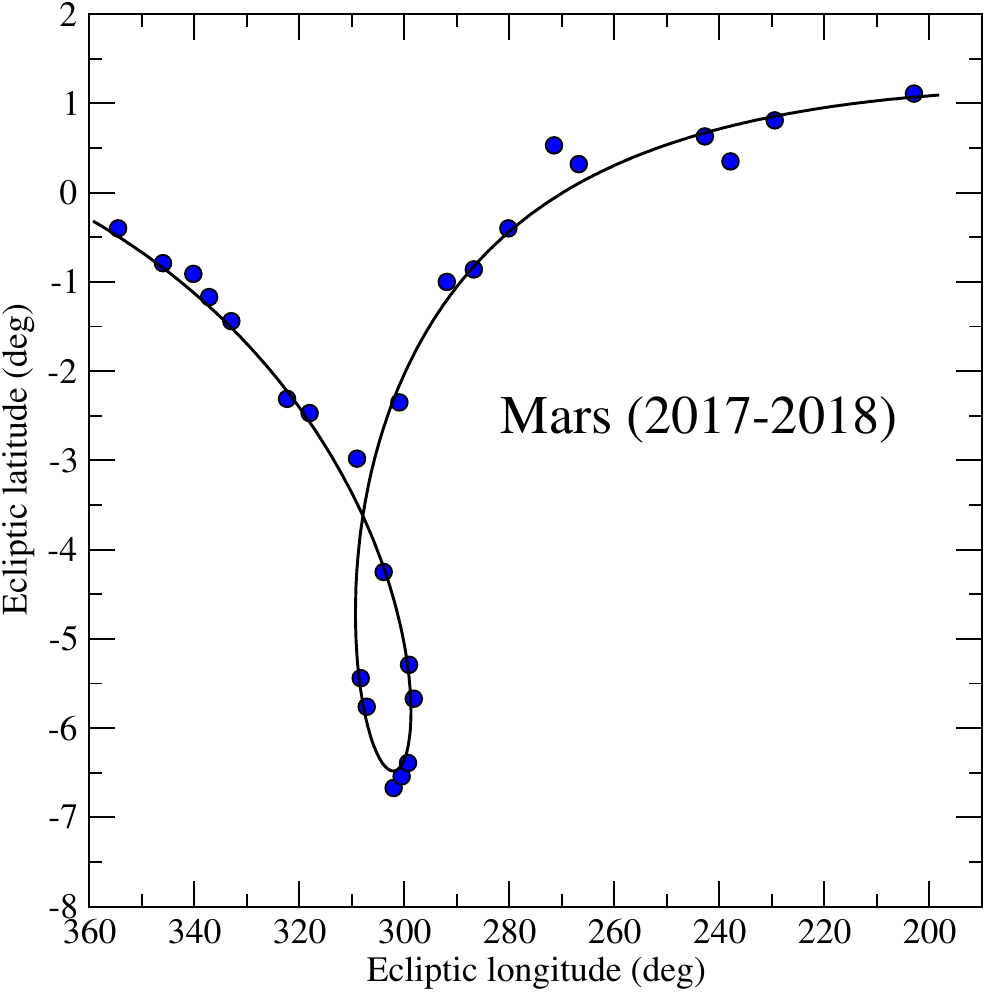} {Krisciunas Fig. \ref{mars_ecl_2018}. 
}

\end{figure}
\begin{figure}
\plotone{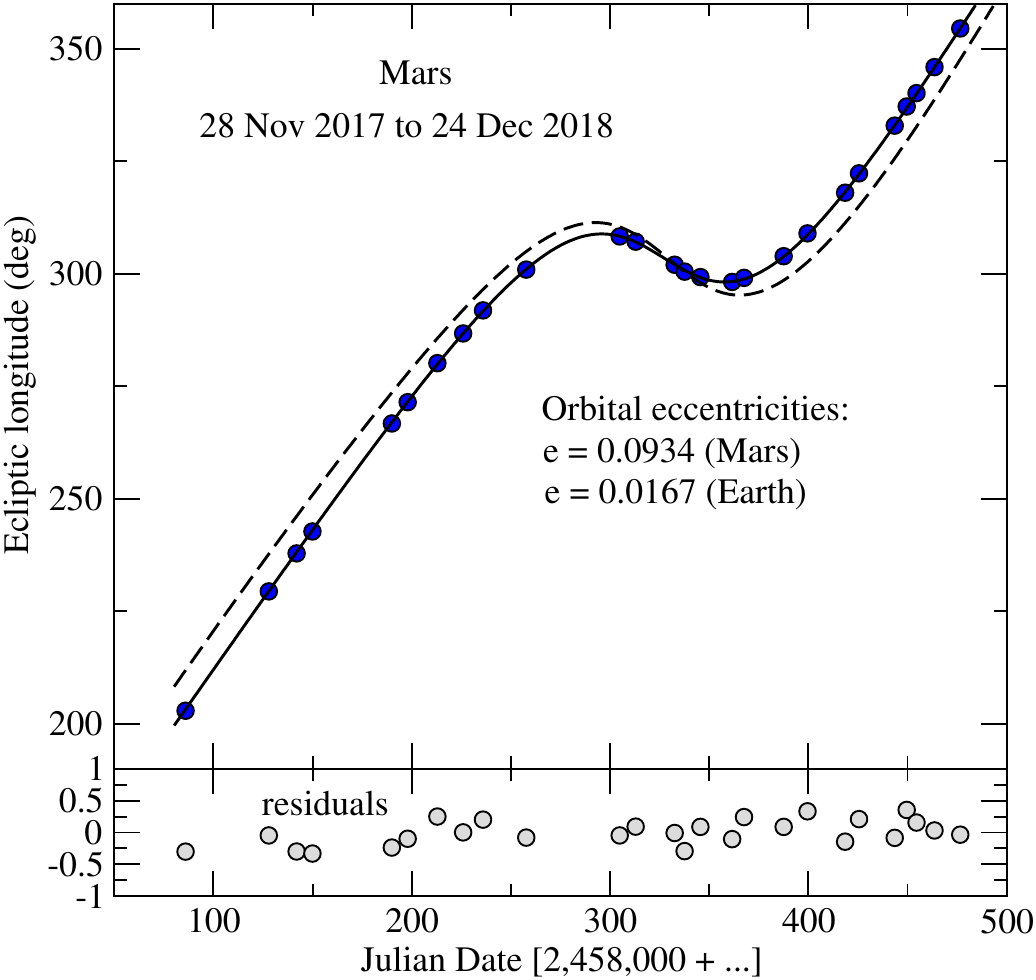} {Krisciunas Fig. \ref{mars_ecc_2018}. 
}
\end{figure}


\begin{thebibliography}{}

\bibitem[Carroll \& Ostlie(2007)]{Car_Ost07} Carroll, B. W., \&
Ostlie, D. A. 2007, {\em An Introduction to Modern Astrophysics}, 2nd ed.,
San Francisco: Pearson/Addison-Wesley

\bibitem[Copernicus(1543)]{Cop1543} Copernicus, N. 1543, {\em On
the Revolutions of the Heavenly Spheres}, translated by A. M. Duncan,
Newton Abbot, London, Vancouver: David \& Charles, 1976, pp. 265-266

\bibitem[Fitzpatrick(2011)]{fitz} Fitzpatrick, R. 2011,
http://farside.ph.utexas.edu/teaching/336k/Newtonhtml/node115.html
(accessed 26 December 2018)

\bibitem[Gingerich(1989)]{Ging89} Gingerich, O. 1989, in {\em The General
History of Astronomy, Vol. 2, Planetary astronomy from the Renaissance
to the rise of astrophysics, Part A: Tycho Brahe to Newton}, R. Taton and
C. Wilson, eds., Cambridge: Cambridge Univ. Press, pp. 54-78

\bibitem[Gingerich(1993)]{Ging93} Gingerich, O. 1993, in {\em The Eye of Heaven:
Ptolemy, Copernicus, Kepler}, New York: American Institute of Physics,
pp. 193-204

\bibitem[Krisciunas(2010)]{Kri10} Krisciunas, K. 2010,
Amer. J. Phys., {\bf 78}, 834

\bibitem[Krisciunas(2016)]{Kri16} Krisciunas, K. 2016, {\em A Guide to Wider
Horizons}, Dubuque, Iowa: Kendall-Hunt, pp. 41-43

\bibitem[Krisciunas et al.(2012)]{Kri_etal12} Krisciunas, K.,
DeBenedictis, E., Steeger, J., Bischoff-Kim, A., Tabak, G., \& Pasricha, K. 2012,
Amer. J. Phys., {\bf 80}, 429

\bibitem[Lahaye(2012)]{Lah12} Lahaye, T. 2012, arXiv:1207.0982

\bibitem[Meeus(1988)]{Meeus88} Meeus, J. 1988, {\em Astronomical Formulae for
Calculators}, 4th ed., Richmond, Virginia: Willman-Bell

\bibitem[Smart(1977)]{Smart77} Smart, W. M. 1977, {\em Textbook on Spherical 
Astronomy}, 6th ed., Cambridge: Cambridge Univ Press

\bibitem[Tholen, Tejfel, \& Cox(2000)]{Tho_etal00} Tholen, D. J., Tejfel, V. G.,
\& Cox, A. N. 2000, in {\em Allen's Astrophysical Quantities}, 4th ed.,
A. N. Cox, ed., New York, Berlin, Heidelberg: Springer, pp. 293-313

\bibitem[Toomer(1984)]{almagest}Toomer, G. J. 1984, {\em Ptolemy's 
Almagest}, Berlin: Springer-Verlag, pp. 254, 284

\bibitem[Toomer(1981)]{hipp} Toomer, G. J. 1981, ``Hipparchus,'' in 
{\em Dictionary of Scientific Biography}, Charles Coulston Gillespie, ed.,
New York: Charles Scribner's Sons, vol. 15, pp. 207-224

\bibitem[van Flandern \& Pulkkinen(1979)]{vanF_Pul79} Van Flandern, T. C.
\& Pulkkinen, K. F. 1979, \apjs, {\bf 41}, 391

\end{thebibliography}
\end{document}